\documentclass[figures,doublecol]{epl2}

\usepackage{amsmath,amssymb,amstext}
\usepackage{color}

\newcommand{\kB}{k_\mathrm{B}}
\newcommand{\Tc}{T_\mathrm{c}}
\newcommand{\Fc}{\mathcal F_\mathrm{C}}
\newcommand{\Phic}{\Phi_\mathrm{C}}
\newcommand{\Dc}{\Delta_\mathrm{C}}
\newcommand{\df}{f_\mathrm{res}}
\newcommand{\dF}{F_\mathrm{res}}
\newcommand{\dZ}{Z_\mathrm{res}}
\newcommand{\F}{F}

\newcommand{\Ls}{L_\perp}
\newcommand{\Lp}{L_\parallel}
\newcommand{\dL}{\delta L}
\newcommand{\dR}{\delta R}
\newcommand{\Leff}{L_\mathrm{eff}}
\newcommand{\Reff}{R_\mathrm{eff}}
\newcommand{\hh}{\tilde{h}}
\newcommand{\pp}{\ddagger}

\newcommand{\zetb}{\bar\zeta}

\begin{document}

\title{Direct simulation of critical Casimir forces}

\author{Hendrik Hobrecht \and Alfred Hucht}
\shortauthor{Hendrik Hobrecht \etal}
\institute{Fakult\"{a}t f\"{u}r Physik und CENIDE, Universit\"{a}t Duisburg-Essen, D-47048 Duisburg, Germany}

\abstract{
We present a new Monte Carlo method to calculate Casimir forces acting on objects in a near-critical fluid, considering the two basic cases of a wall and a sphere embedded in a two-dimensional Ising medium. 
During the simulation, the objects are moved through the system with appropriate statistical weights, and consequently are attracted or repelled from the system boundaries depending on the boundary conditions.
The distribution function of the object position is utilized to obtain the residual free energy, or Casimir potential, of the configuration as well as  the corresponding Casimir force. 
The results are in perfect agreement with known exact results.
The method can easily be generalized to more complicated geometries, to higher dimensions, and also to colloidal suspensions with many particles.
}

\maketitle

\section{Introduction}

Casimir forces appear whenever a medium with long-range fluctuations is confined to a restricted geometry.
They were predicted by H.~B.~G.~Casimir in 1948 as attractive forces between two conducting plates in vacuum, caused by fluctuations of the electromagnetic field \cite{Casimir48}. 
Analogously, in thermodynamic systems the so called critical Casimir forces appear near a continuous phase transition, induced by the long-range correlated fluctuations of the order parameter \cite{FisherDeGennes1978}. 
One of the experimental evidences for this effect is the change of thickness of critical liquid films, which was measured by Garcia and Chan for $^{4}$He films \cite{GarciaChan1999} close to the $\lambda$-point and near the $^{3}$He-$^{4}$He tricritical point \cite{GarciaChan2002}.
The Ising class was experimentally studied by Fukuto \emph{et al.} \cite{Fukuto2005} with binary wetting films. 
Hertlein \emph{et al.} \cite{HertleinHeldenGambassiDietrichBechinger08} were able to measure the critical Casimir force between a single colloidal sphere and a flat surface in a binary liquid due to the measurement of the distribution function of the distance between the particle and the surface. 
Recently, Casimir forces were used by Nguyen \emph{et al.} \cite{Nguyen2013} to control the critical aggregation of colloids in binary liquids.
A theoretical description of the critical Casimir force between two spheres as well as between a sphere and a surface in arbitrary dimension was given in \cite{BurkhardtEisenriegler1995}, while two arbitrarily shaped objects in two dimensions were recently discussed in \cite{Bimonte2013}. 
In the last few years, Monte Carlo (MC) simulations were used, e.~g., by Hucht \cite{Hucht2007} and Vasilyev \emph{et al.} \cite{Vasilyev2007} to examine critical Casimir forces and especially the form of the according scaling functions, which is -- in most cases -- not possible analytically. 
Recently, a new MC algorithm to compute the critical Casimir force between a sphere and a surface was introduced by Hasenbusch \cite{Hasenbusch2013}. 
While this method allows a very accurate determination of Casimir forces, practically it is limited to one object degree of freedom.

The Casimir force per unit area $A=\Lp^{d-1}$ of a system in film geometry
$\Lp^{d-1}\times \Ls$ with periodic boundary conditions (BC) in $\Lp$-direction at reduced temperature $t=T/\Tc-1$ near the critical temperature $\Tc$ is defined as
\begin{align}\label{eq:Fc}
 \beta\Fc (t,\Ls,\Lp) \equiv
  -\frac{1}{A}\frac{\partial}{\partial \Ls}\dF(t,\Ls,\Lp),
\end{align}
with $\beta=1/\kB T$.
The total residual free energy $\dF$, also known as \emph{Casimir potential} in the context of colloidal particle aggregation \cite{HertleinHeldenGambassiDietrichBechinger08,Nguyen2013,Soyka2008}, is given by
\begin{align}
 \dF(t,\Ls,\Lp) \equiv
 \F(t,\Ls,\Lp)-V f_{\mathrm{b}}(t)-2A f_{\mathrm{s}}(t),
\end{align}
with total free energy $\F(t,\Ls,\Lp)$, bulk free energy density $f_{\mathrm{b}}(t)$, surface free energy per area $f_{\mathrm{s}}(t)$ and Volume $V=\Lp A$.
All free energies are measured in units of $\kB T$.
 
Following Fisher and de Gennes \cite{FisherDeGennes1978}, the Casimir force $\Fc$ fulfills the scaling \textit{ansatz}\footnote{
 Throughout this work, the symbol $\simeq$ means asymptotically equal
 in the respective limit, e.g., $f(L)\simeq
 g(L)\Leftrightarrow\lim_{L\rightarrow\infty}f(L)/g(L)=1$. 
}
\begin{align} \label{eq:casforce1}
 \beta\Fc(t,\Ls,\Lp)\simeq \Ls^{-d}\vartheta(x,\rho)
\end{align} 
near $\Tc$ and for large $\Ls$, $\Lp$.
The universal finite-size scaling function $\vartheta(x,\rho)$ depends on the temperature scaling variable $x$ and aspect ratio $\rho$,
\begin{align}
 x \equiv t\left(\frac{\Ls}{\xi_{0}^{+}}\right)^\frac{1}{\nu},\quad
 \rho \equiv \frac{\Ls}{\Lp},
\end{align}
with critical exponent $\nu$ and correlation length amplitude $\xi_{0}^{+}$ defined by $\xi(t>0)\simeq\xi_{0}^{+} t^{-\nu}$.

An analogous \textit{ansatz} can be made for the residual free energy $\dF$ and the residual free energy per surface area $\df\equiv\dF/A$,%
\begin{subequations}\label{eq:PhiTheta(x,rho)}
\begin{align}
 \df(t,\Ls,\Lp)&\simeq \Ls^{-(d-1)}\Theta(x,\rho)\label{eq:Theta(x,rho)},\\
 \dF(t,\Ls,\Lp)&\simeq \Phic(x,\rho),\label{eq:Phi(x,rho)}
\end{align}
\end{subequations}
with universal finite-size scaling functions $\Phic$ and $\Theta$ fulfilling
\begin{align}\label{eq:Phi_of_Theta}
 \Phic(x,\rho)=\rho^{1-d}\Theta(x,\rho),
\end{align}
see Ref.~\cite{HuchtGruenebergSchmidt2011} for details\footnote{
The Casimir potential obeys $\Phic(x,\rho)=\Theta(x \rho^{\frac{1}{d\nu}-\frac{1}{\nu}},\rho)$, with $\Theta$ from Eq.~(12) in Ref.~\cite{HuchtGruenebergSchmidt2011}.
}.
On the other hand, $\vartheta(x,\rho)$ and $\Theta(x,\rho)$ satisfy the scaling relation
\begin{align}
 \vartheta(x,\rho)=(d-1)\Theta(x,\rho)-\frac{x}{\nu}\frac{\partial \Theta(x,\rho)}{\partial x}-\rho\frac{\partial \Theta(x,\rho)}{\partial\rho}.
\end{align}
At the critical temperature $t=0$ and for $\rho \to 0$ the Casimir force simplifies to
\begin{align}
 \beta\Fc(0,\Ls,\infty)\simeq \Ls^{-d} (d-1) \Dc,
\end{align}
with the universal Casimir amplitude in film geometry, $\Dc\equiv\Theta(0,0)$.

\section{The mobile wall} 
Common theoretical methods \cite{Hucht2007,Vasilyev2007,DieGruHaHuRuSch12,Hasenbusch2013} to obtain critical Casimir forces involve the calculation of the free or internal energy of certain fixed geometries, e.g., films with constant thickness $\Ls$, combined with a derivative of the energy with respect to $\Ls$.
These methods, though quite successful, imply a few drawbacks: 1) they can only be used for simple geometries as the slab geometry or the surface-sphere geometry; 2) they usually require knowledge of the corresponding bulk energy; and 3) it is necessary to perform a thermodynamic integration in order to get the free energy.

In this work we will choose a completely different way and determine the residual free energy and the Casimir force dynamically within systems with geometrical degrees of freedom. Therefore, this method is very similar to the experiments on colloidal particles as performed by Hertlein \emph{et al.} \cite{HertleinHeldenGambassiDietrichBechinger08}.

\begin{figure}[t]
	\centerline{\includegraphics[width=0.48\textwidth,clip,trim=1.8cm 14.8cm 4cm 3cm]{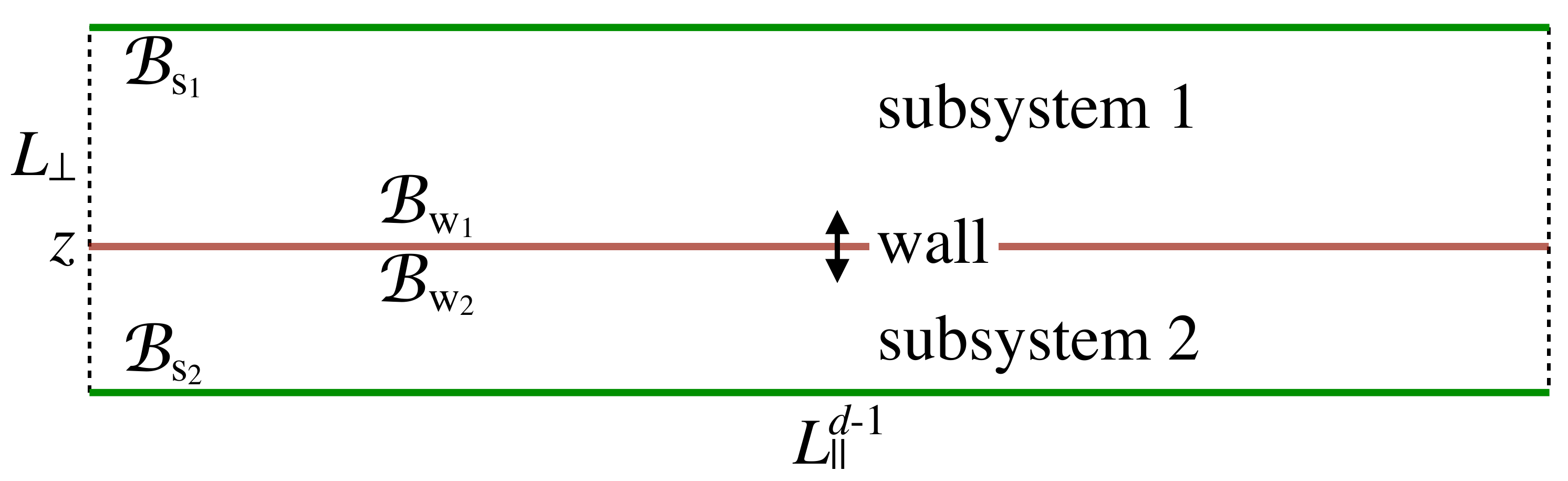}}
	\caption{(color online) Sketch of the system with a mobile wall at position $z$.}
	\label{fig:systemWall}
\end{figure}

The system under study is a $d$-dimensional Ising model with spin variables $\sigma=\pm 1$, nearest-neighbor couplings $J$ and Hamiltonian
\begin{align}
  \mathcal H = -J \sum_{\langle ij\rangle}\sigma_i \sigma_j
  -\sum_{\sigma_i\in\mathcal B_\mu} b_\mu \sigma_i.
\end{align}
We assume periodic BCs in the $\parallel$ directions, while the BCs in the $\perp$ direction are modeled using the boundary fields $b_\mu=\{\pm 1,0\}$ (for fixed or open BCs), acting on the top ($\mu=\mathrm s_1$) and bottom ($\mu=\mathrm s_2$) boundary spins $\sigma_i\in\mathcal B_\mu$. 
Within the spin medium we place a parallel intermediate wall at position $0<z<\Ls$, such that spins above ($\mu=\mathrm w_1$) and below the wall ($\mu=\mathrm w_2$) also become boundary spins, see Fig.~\ref{fig:systemWall}.
The wall is mobile and, depending on the combination of BCs, is attracted or repelled from the system boundaries\footnote{A similar geometry has been investigated in the framework of QED Casimir forces \cite{Hertzberg2005}.}.

The partition function of the system with wall reads
\begin{align}
Z&=\sum_{z}\sum_{\{s\}} e^{-\beta\mathcal H(\Ls,\Lp;z)}
  =\sum_{z} e^{-\F(t,\Ls,\Lp;z)}\nonumber\\
 &=Z_\mathrm{bs} \sum_{z} e^{-\dF(t,\Ls,\Lp;z)}
 \equiv Z_\mathrm{bs} \dZ.
\end{align}
The constant $Z_\mathrm{bs}$ contains the bulk and surface terms, which are independent of the wall position and cancel out in the following. 
As a consequence, the probability distribution function of the wall position $z$ becomes
\begin{subequations}\label{eq:hht}
\begin{align}\label{eq:h(t,Ls,Lp,z)}
 h(t,\Ls,\Lp;z)=\frac{1}{\dZ} e^{-\dF(t,\Ls,\Lp;z)}
\end{align}
and fulfills the \textit{ansatz} $h(t,\Ls,\Lp;z)\simeq\hh(x,\rho;\zeta)$, with scaling function
\begin{align}\label{eq:h(x,rho,zeta)}
 \hh(x,\rho;\zeta)=\frac{1}{\Sigma_\mathrm{C}}e^{-\Phic(x,\rho;\zeta)}.
\end{align}
\end{subequations}
Here we introduced the reduced wall position $\zeta=z/\Ls$, whereas the normalization $\Sigma_\mathrm{C}$ ensures that $\int_0^1 \mathrm d\zeta\,\hh(\zeta)=1$.

As the wall separates the system into two non-inter{-}acting subsystems with thickness $z$ and $\Ls{-}z$, we can calculate the asymptotic form of $\Phic(x,\rho;\zeta)=\rho^{1-d} \Theta(x,\rho;\zeta)$ as the sum of two contributions, with
\begin{align}\label{eq:Theta(x,rho,z)}
\Theta(x,\rho;\zeta)
 =\frac{\Theta^{(1)}(\zeta^\frac{1}{\nu} x,\zeta\rho)}{\vphantom{\zetb}{\zeta}^{d-1}}
 +\frac{\Theta^{(2)}(\zetb^\frac{1}{\nu} x,\zetb\rho)}{\zetb^{d-1}},
\end{align}
where $\zetb\equiv 1-\zeta$, and $\Theta^{(1,2)}$ denote the residual free energy scaling functions, Eq.~\eqref{eq:Theta(x,rho)}, of subsystems $1$ and $2$.

In analogy to Eq.~\eqref{eq:Fc}, the Casimir force per area acting on the wall is given by the derivative of the residual free energy per area with respect to $z$,
\begin{align}
 \beta\Fc(t,\Ls,\Lp;z) = -\frac{\partial}{\partial z}\df(t,\Ls,\Lp;z),
\end{align}
with corresponding scaling form
\begin{align}\label{eq:theta(x,rho,z)}
\vartheta(x,\rho;\zeta)
=\frac{\vartheta^{(1)}(\zeta^\frac{1}{\nu} x,\zeta\rho)}{\zeta^{d}}
-\frac{\vartheta^{(2)}(\zetb^\frac{1}{\nu} x,\zetb\rho)}{\zetb^{d}}.
\end{align}

We first focus on the critical point $T=\Tc$, where $x=0$. 
In the limit of a thin two-dimensional Ising film, $d=2$, $\nu=1$, and $\rho \ll 1$, and equal subsystem boundaries considered in this work, Eqs.~\eqref{eq:Theta(x,rho,z)} and \eqref{eq:theta(x,rho,z)} simplify to
\begin{subequations}
\begin{align}
 \Theta   (0,0;\zeta) &= \Dc\left(\zeta^{-1}+\zetb^{-1}\right),\\
 \vartheta(0,0;\zeta) &= \Dc\left(\zeta^{-2}-\zetb^{-2}\right),
\end{align}
\end{subequations}
with universal Casimir amplitude for antisymmetric boundary
conditions $\Dc=23\pi/48$ \cite{EvansStecki1994}.
These predictions are now checked within MC simulations.

\section{MC method}

In the MC simulations we consider a ferromagnetic Ising model with spin variables $s=\pm 1$, nearest neighbor coupling $J=1$ and Hamiltonian
\begin{align}
 \mathcal{H}=-\sum_{x=0}^{\Lp-1}\sum_{y=0}^{\vphantom{\Lp}\Ls+1}
 s_{x,y}(s_{x+1,y}+s_{x,y+1})
\end{align}
under a single-spin-flip algorithm. The intermediate wall at position $z$ as well as the BCs at the surfaces are realized via lines of fixed spins according to $s_{x,0}=s_{x,\Ls+2}=b_\mathrm{s}=1$ and $s_{x,z}=b_\mathrm{w}=-1$, leading to $\Ls$ spin degrees of freedom per surface area. 
The intrinsic time scale is one MC sweep, i.e., when on average all free spins had the chance to be updated once.

The wall position is updated once every MC sweep.
The direction of the motion is chosen randomly with probability $1/2$ for a step up or down, and the wall interchanges its position with the next row of spins in the chosen direction under spin conservation.
Therefore the couplings in $x$-direction are not changed, and the energy difference is given by
\begin{align}\label{eq:DeltaE}
 \Delta E_\mu=-\sum_{x=0}^{\Lp-1}(s_{x,z+\mu}-s_{x,z})(s_{x,z+2\mu}-s_{x,z-\mu}),
\end{align}
with $\mu=\pm 1$ for up/down steps, respectively.
With this energy difference, the wall move is accepted or rejected via the usual Metropolis algorithm. 

Note that in this method we do not require a separation of time scales to obtain equilibrium configurations, with spin dynamics much faster than object dynamics, because the object position update fulfills detailed balance. 
Specifically, a movement of the wall via Eq.~\eqref{eq:DeltaE} is only accepted if the spin system is in equilibrium for the new wall position. 
This is different from a realistic dynamics like, e.g., Brownian motion, where a separation of time scales is indeed necessary.

\begin{figure}[t]
	\centerline{\includegraphics[width=0.48\textwidth]{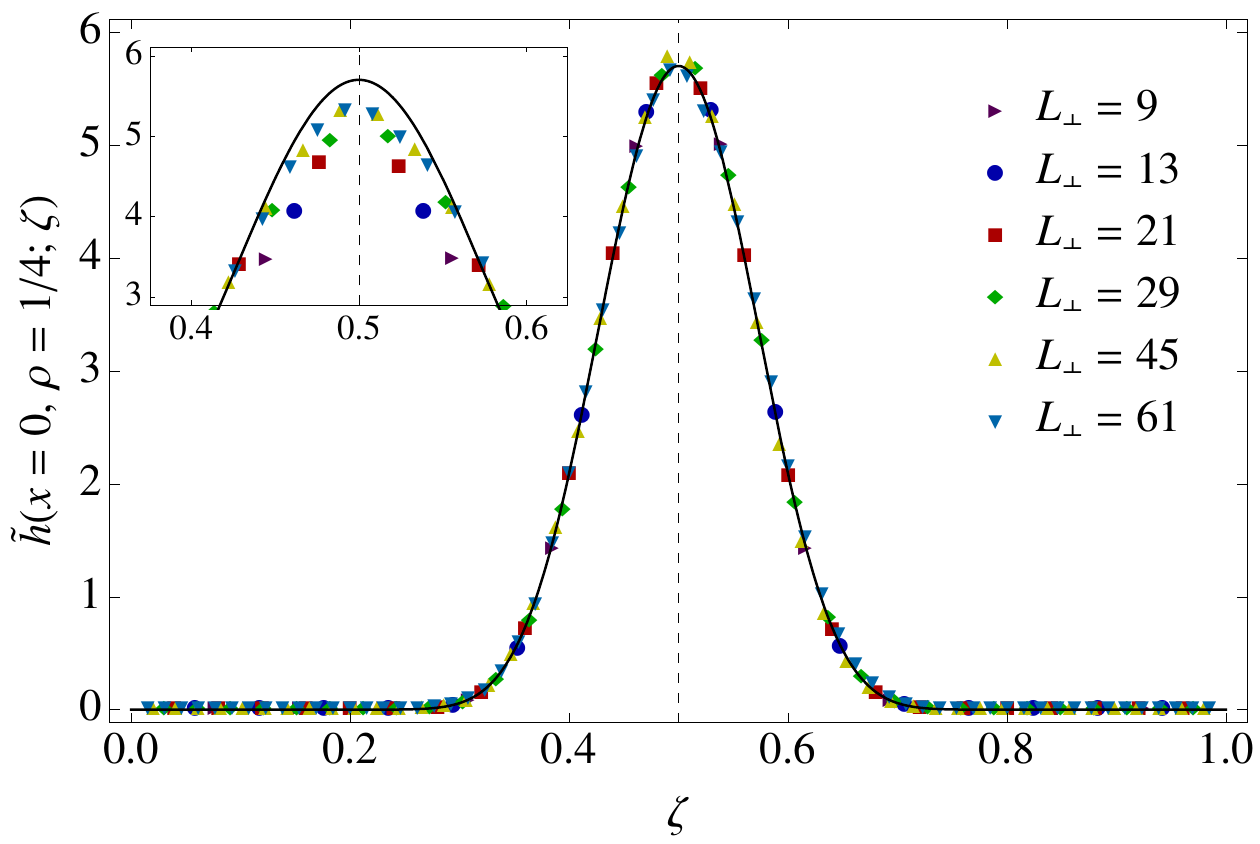}}
	\caption{(color online) The distribution function $\hh(x,\rho;\zeta)$, Eq.\eqref{eq:h(x,rho,zeta)}, of the rescaled wall position $\zeta$ for temperature $x=0$ and aspect ratio $\rho=1/4$ is shown as a solid line, together with the rescaled histograms of the MC simulations with thickness correction $\dL=0.87(10)$. The inset shows the uncorrected data.}
	\label{fig:histo_0}
\end{figure}

From the data of the MC simulations with $10^6$ MC sweeps per system size the distribution function, Eq.~\eqref{eq:hht}, is calculated as a histogram of the time series of the wall position $z$. 
The examined system sizes are $\Ls=\{9,13,21,29,45,61\}$, whereas $z=\{1,2,\dots,\Ls+1\}$ due to the fixed boundaries at $z=0$ and $z=\Ls+2$. 
The simulations are run at the critical temperature $x=0$ and with constant aspect ratio $\rho=1/4$, which gives the subsystems an average aspect ratio of $\rho^{(1,2)}=1/8$. 
This aspect ratio satisfies the condition $\rho\ll 1$ \cite{HuchtGruenebergSchmidt2011} and we can use the argument $\rho=0$ in the corresponding scaling functions. 
The results are shown in the inset of Fig.~\ref{fig:histo_0} and display non-negligible corrections to scaling from the discrete lattice.

In the following we will demonstrate that these corrections to scaling can be largely eliminated by a thickness correction $\dL$ per surface to the system length $\Ls$, leading to the effective thickness \cite{DieGruHaHuRuSch12}
%
\begin{align}
 \Leff=\Ls+4\,\dL
\end{align}
and effective wall position
\begin{align}
 z_\mathrm{eff} = z - 1 + 2\,\dL.
\end{align}
%
With these definitions, the reduced wall position reads $\zeta = z_\mathrm{eff}/\Leff$, and Eqs.~\eqref{eq:PhiTheta(x,rho)} become
\begin{subequations}\label{eq:PhiTheta_eff(x,rho)}
\begin{align}
   \df(t,\Ls,\Lp)&\simeq \Leff^{-(d-1)}\Theta(x,\rho),\label{eq:Theta_eff(x,rho)}\\
   \dF(t,\Ls,\Lp)&\simeq \left(\frac{\Leff}{\Ls}\right)^{-2(d-1)}\Phic(x,\rho)\label{eq:Phi_eff(x,rho)}.
\end{align}
\end{subequations}
Note that both $x$ and $\rho$ have to be calculated with $\Leff$ instead of $\Ls$, leading to the factor of two in the exponent of Eq.~\eqref{eq:Phi_eff(x,rho)}, see Eq.~\eqref{eq:Phi_of_Theta}.
For $\dL=0.87(10)$ the data sets of the different system sizes nicely collapse onto a single curve as shown in Fig.~\ref{fig:histo_0}. 

The finite-$L$ corrections to the Casimir amplitude $\Dc$ for the 2d Ising case in slab geometry $\rho\to0$ and with symmetric BCs can be calculated exactly from Eqs.~(2.5) of Ref.~\cite{EvansStecki1994}. Substituting $\varphi\to\phi/\Ls$ in the integrand, expanding around $\Ls=\infty$ and integrating by terms over $\phi\in[0,\infty]$ gives the large-$L$ expansion of the residual free energy $\df^{(\pp)}(0,\Ls,\infty)$.
The antisymmetric case $\df^{(\pm)}(0,\Ls,\infty)$ is obtained from Eq.~(3.1ff) of Ref.~\cite{EvansStecki1994}.
Inserting these results into Eq.~\eqref{eq:Theta_eff(x,rho)} we derive the effective thickness
\begin{align}\label{eq:Leff}
 \Leff^{(\wp)} = \Ls + 2\dL + \frac{a^{(\wp)}}{\Ls} \left(1-\frac{1+\sqrt 2}{\Ls}\right)+\mathcal O(\Ls^{-3}),
\end{align}
with $2\dL=1+1/\sqrt 2$ for two surfaces. 
Hence the first-order correction $\dL$ turns out to be independent of the BC symmetry, whereas the next-order correction reads $a^{(\pp)}=-7\pi^2/480$ for symmetric BC and $a^{(\pm)}=247\pi^2/11040$ for antisymmetric BC.
Note that in the derivation of the correction amplitude $a^{(\pm)}$ we identified two typos in Eq.~(3.4) of Ref.~\cite{EvansStecki1994}, which correctly reads
\begin{align}
 \gamma_1 = \frac{\pi}{2M} - \frac{\sqrt 2\pi}{4M^2} + \frac{\pi}{4M^3}\left(1-\frac{\pi^2}{24}\right) \nonumber\\
{} - \frac{\sqrt 2\pi}{8M^4}\left(1-\frac{\pi^2}{6}\right)+\mathcal O(M^{-5})
\end{align}
with $M=\Ls+1$. 
The $M^{-4}$ term is required for the $\Ls^{-2}$ correction in Eq.~\eqref{eq:Leff}.

\begin{figure}[t]
	\centerline{\includegraphics[width=0.48\textwidth]{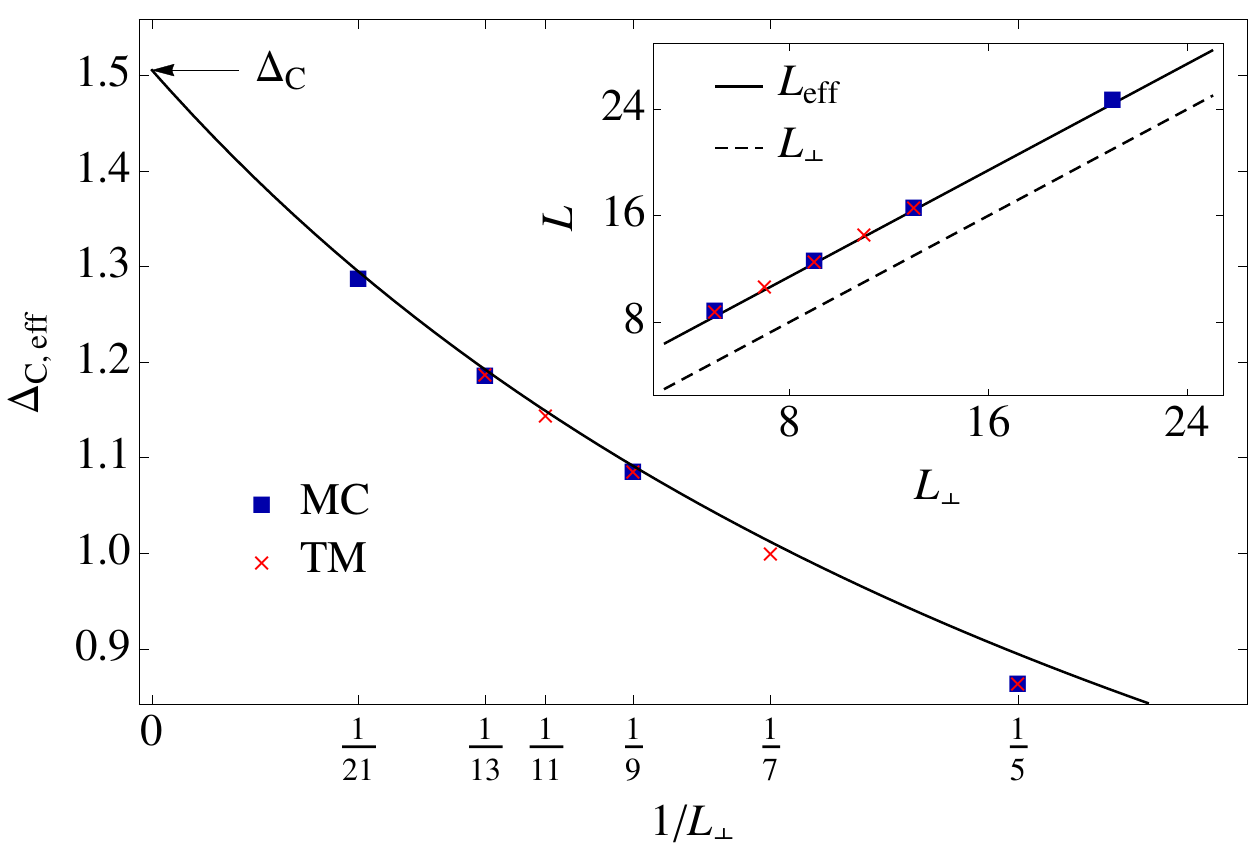}}
	\caption{(color online)
	Effective Casimir amplitude over $1/\Ls$ as determined from the numerics.
	The crosses are from the TM calculations, the other symbols are from MC simulations. 
	The solid line is $\Delta_{\mathrm{C,eff}}(\Ls)$ from Eq.~\eqref{eq:Dc_eff}, while the inset shows $\Leff(\Ls)=\Dc/\df(0,\Ls,\Lp)$.
	Even for small $\Ls$ the higher-order corrections to $\Leff$ are small and can be neglected.
        }
	\label{fig:casampc}
\end{figure}

\begin{figure}[t]
\includegraphics[width=0.48\textwidth]{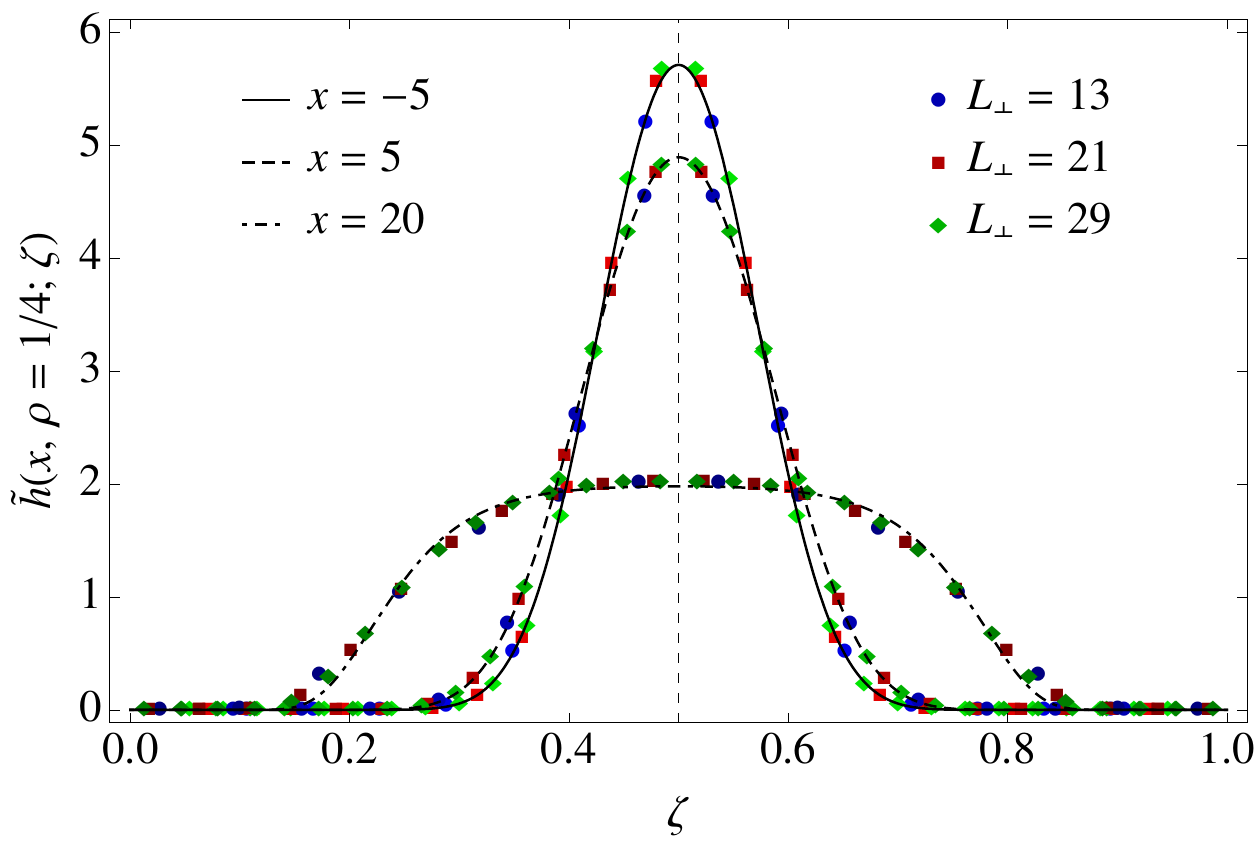}%
	\caption{(color online) The distribution function $\tilde{h}(x,\rho;\zeta)$ of the rescaled wall position $\zeta$ for aspect ratio $\rho=1/4$ is shown for three values $x=\{-5,5,20\}$ of the temperature scaling variable.
	 The symbols are results from the MC simulations with adapted $\dL$ (see text), the black lines are theoretical predictions from Eq.\eqref{eq:h(x,rho,zeta)} and Eq.\eqref{eq:Theta_2d_ex}.
        }
	\label{fig:histo_pm5}
\end{figure}

Since our MC simulations are in very good agreement with the predicted distribution function, we now use them to numerically determine the Casimir amplitude $\Dc$ and its finite-size corrections.
Therefore we plot the effective Casimir amplitude 
\begin{align}\label{eq:Dc_eff}
 \Delta_\mathrm{C,eff}(\Ls) \equiv \Ls\df(0,\Ls,\Lp) \simeq \Dc\frac{\Ls}{\Leff}
\end{align}
over $1/\Ls$,
see Eqs.~\eqref{eq:Theta(x,rho)} and \eqref{eq:Theta_eff(x,rho)},
and find $\Dc=1.50(3)$ and $\dL=0.9(1)$ in good agreement with the exact $\rho\to0$ results $\Dc=1.5053\dots$ and $\dL=0.85355\dots$.
Additionally, we calculated the numerically exact residual free energy of one film using standard numerical transfer matrix (TM) methods with corresponding BCs and given thickness and aspect ratio.
The total residual free energy is the sum of two such systems, and
varying the wall position is equivalent to a corresponding change in the thickness and aspect ratio of the subsystems.
From the residual free energy we calculated the distribution function for $\Ls=\{5,7,9,11,13\}$ and $\rho=1/4$, rescaled and normalized it and subsequently used Eq.~\eqref{eq:hht} to fit $\Dc$.
Figure~\ref{fig:casampc} shows the data from the TM calculations in comparison with the MC simulations together with $\Delta_\mathrm{C,eff}(\Ls)$ from Eq.~\eqref{eq:Dc_eff}.
The inset of Fig.~\ref{fig:casampc}, showing the numerically calculated $\Leff(\Ls)=\Dc/\df(0,\Ls,\Lp)$, demonstrates that higher-order corrections to $\Leff$ can indeed be safely neglected.

We now turn to temperatures $T\neq\Tc$.
In the case of the two-dimensional Ising model, where the correlation length  exponent $\nu=1$, the scaling variables $x^{(1,2)}$ of the subsystems are linear in $\zeta$ and given by
\begin{align}
\label{eq:xeff}
 x^{(1)}=\zeta x,
 &&
 x^{(2)}=\zetb x,
\end{align}
see Eq.~\eqref{eq:Theta(x,rho,z)}.
The global scaling variable simply becomes
\begin{align}
 x=x^{(1)}+x^{(2)},
\end{align}
since both subsystems have the same temperature.
For $\rho\to0$ the scaling function of the system reads
\begin{align}\label{eq:Theta_2d_ex}
\Theta(x,0;\zeta)
=\frac{\Theta^{(\pm)}(\zeta x)}{\zeta}
+\frac{\Theta^{(\pm)}(\zetb x)}{\zetb},
\end{align}
where the residual free energy scaling function $\Theta^{(\pm)}(x)$ for a strip with $\pm$ BC is known exactly from the work of Evans and Stecki \cite{EvansStecki1994}. 

The simulations were run for $\Ls=\{13,21,29\}$ and $x=\{-5,5,20\}$, where we used the exactly known correlation length amplitude $\xi_{0}^{+}=[2\ln(1+\sqrt{2})]^{-1}$. 
As explained above we calculated $x$ and $\rho$ using $\Leff$ instead of $\Ls$, using $\dL=\{0.875(50),0.75(5),0.20(5)\}$ for $x=\{-5,5,20\}$.
The results are displayed as symbols in Fig.~\ref{fig:histo_pm5} together with the  distribution functions derived from Eq.~\eqref{eq:Theta_2d_ex} and again show a convincing data collapse. 
For large values of $x$ the Casimir force becomes small in the center of the system, leading to a flat histogram in this region. 
This is shown exemplarily for $x=20$.

\section{The mobile sphere}

\begin{figure}[b!]
	\centerline{\includegraphics[width=0.48\textwidth,clip,trim=1.8cm 4.2cm 4cm 13.8cm]{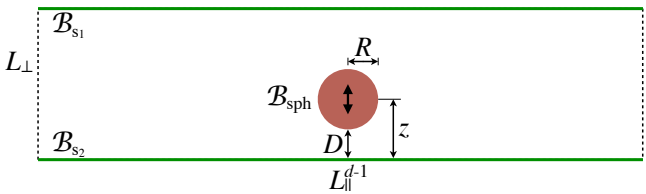}}
	\caption{(color online) Sketch of the system with a mobile sphere at position $z$.}
	\label{fig:systemSphere}
\end{figure}

Finally we present first results for a mobile sphere with radius $R$ confined in a $d$-dimensional system with thickness $\Ls$ at criticality $T=\Tc$. 
Due to the periodic BCs in $||$ direction, we can restrict the motion of the sphere to perpendicular moves without loss of generality.
Then the sphere position $z$ is restricted to $R<z<\Ls-R$, and we define the reduced sphere coordinate $\zeta=D/(\Ls-2R)$, with sphere-surface distance $D=z-R$, see Fig.~\ref{fig:systemSphere}.

Analogously to Eq.~\eqref{eq:hht}, the sphere-position distribution is given by 
\begin{subequations}\label{eq:hht_sph}
\begin{align}
h(0,\Ls,\Lp,R;z)&=\frac{1}{\dZ} e^{-\dF(0,\Ls,\Lp,R;z)} \\
\simeq \hh(0,\rho,r;\zeta)&=\frac{1}{\Sigma_\mathrm{C}} e^{-{\Phic}(0,\rho,r;\zeta)}
\end{align}
\end{subequations}
with residual free energy or Casimir potential $\dF$, associated scaling function $\Phic$, and the reduced sphere radius $r=R/\Ls$. 
Since the system no longer divides up into two subsystems, the free energy does not separate as in the wall case.
Furthermore, we do not know of any conformal mapping that can be used to exactly calculate the Casimir force in this effective three body setup.
Nevertheless, we can compare the simulations with known results for small distances $\zeta$.

In two-dimensional systems, the critical residual free energy at small $\zeta$ is given by \cite{BurkhardtEisenriegler1995}%
\begin{subequations}\label{eq:dFDc_sph}
\begin{align}\label{eq:dF_sph}
 \Phic(0,\rho,r;\zeta) \simeq \Dc^<(r)\,\zeta^{-1/2},
\end{align}
with amplitude
\begin{align}\label{eq:Dc_sph}
 \Dc^<(r) &= 2\pi \Dc\,(2 r^{-1} - 4)^{-1/2}.
\end{align}
\end{subequations}
We simulated an Ising system with antisymmetric sphere-surface BCs and aspect ratio $\rho=1/4$ just as in the wall case, and considered a sphere with fixed reduced radius $r=1/16$.
Note that the crossover to the long distance behavior ($D\gg R$) of Eq.~\eqref{eq:dF_sph} predicted in Ref.~\cite{BurkhardtEisenriegler1995} can only occur for a reduced sphere radius $r\lesssim 0.006$ in the considered film geometry\footnote{This value is obtained by equating the two expressions given in Eq.~(17) of Ref.~\cite{BurkhardtEisenriegler1995} at $\zeta=1/2$.}.

\begin{figure}[t]
	\centerline{\includegraphics[width=0.48\textwidth]{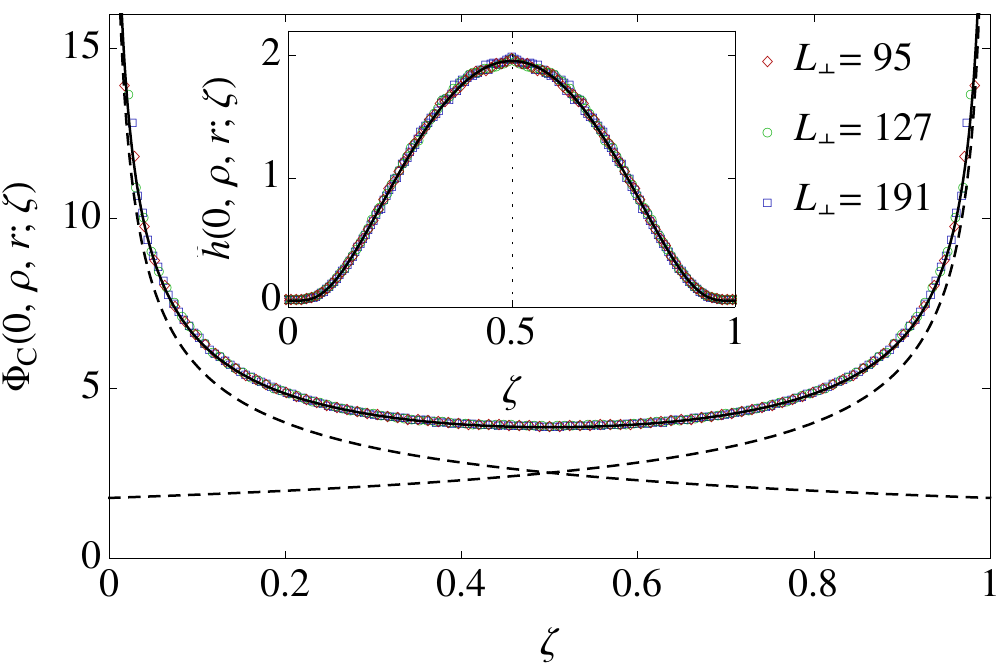}}
	\caption{(color online) Critical Casimir potential $\Phic(0,\rho,r;\zeta)$ of the sphere geometry for aspect ratio $\rho=1/4$ and reduced sphere radius $r=1/16$. 
	The exact asymptotics, Eq.~\eqref{eq:dFDc_sph}, and the fit function, Eq.~\eqref{eq:PhiC_fit}, are shown as dashed and solid lines, respectively. 
	The inset shows the distribution function of the sphere position $\zeta$, Eq.~\eqref{eq:hht_sph}.}
	\label{fig:swg}
\end{figure}

Figure~\ref{fig:swg} shows the reduced residual free energy of the system for $\Ls=\{95,127,191\}$ from the MC simulations, obtained as logarithm of the histogram shown in the inset.
We again use an effective thickness $\Leff=\Ls+2\dL$ as well as an effective sphere radius $\Reff=R+\dR$ to eliminate leading scaling corrections \cite{Hasenbusch2013}. With $\dL$ from Eq.~\eqref{eq:Leff} we find $\dR=1.1(5)$, slightly larger than $\dL$.

As the sphere is realized using fixed spins having a distance smaller than $R$ to the center of the sphere at position $z$, we expect additional corrections due to the rugged sphere surface especially for small $R$. 
However, the satisfactory data collapse especially for small $\zeta$ emphasizes that these additional discretization effects are negligible for our considered system sizes, where $6 \leq R \leq 12$.

Figure~\ref{fig:swg} also includes the exact curves for the sphere close to one of the two surfaces, Eqs.~\eqref{eq:dFDc_sph}, as well as a fit to the data of the form
\begin{align}\label{eq:PhiC_fit}
 \Phic(0,\rho,r;\zeta)
 \simeq \frac{\Dc^<(r)}{\sqrt{\zeta(1-\zeta)}}+a \ln\left[\zeta(1-\zeta)\right].
\end{align}
Here we made a product \textit{ansatz} with logarithmic corrections and fit parameters $a=-0.21(1)$, $\Sigma_\mathrm{C}=0.0107(1)$, leading to a satisfactory approximation. 
We note that the present geometry leads to substantial deviations from the small $\zeta$ results. 
Additional simulations for smaller aspect ratios $\rho$ suggest that $a$ is approximately proportional to $\rho$.

\section{Conclusions}

In summary, we presented a new method to directly simulate critical Casimir forces acting on objects in a near-critical fluid using Monte Carlo methods.
We considered the two cases of a wall and a sphere, moving within a two-dimensional Ising medium. 
Depending on their boundary conditions, the objects are attracted or repelled from the system boundaries.
Our analysis is based on the position distribution function of those objects, which is independent of any bulk and surface contributions which usually have to be taken into account for a correct determination of the Casimir force. 
From this distribution function the residual free energy, or Casimir potential, and the Casimir force can easily be calculated.
For the wall case we compared the Monte Carlo results near the critical point $\Tc$ both with exact results and with numerical transfer matrix calculations, verifying the validity of the method. 
We verified that scaling corrections from symmetry breaking boundaries can be largely eliminated using an effective length $\Leff$ \cite{DieGruHaHuRuSch12,Hasenbusch2013}.
The results for the sphere geometry are consistent with the theoretical predictions available for small sphere-surface distances.

The presented method can easily be generalized \cite{HobrechtHucht2014b} to more complicated geometries, to systems with multiple objects like colloidal suspensions \cite{Soyka2008,Veen2012}
or inclusions in biological membranes \cite{Machta2012},
as well as to higher dimensions. 
An efficient cluster version of the algorithm is under development and will be presented elsewhere.

\begin{acknowledgments}
We thank Felix M.~Schmidt and Martin Hasenbusch for valuable discussions.
\end{acknowledgments}

\bibliographystyle{../../../TeX/bst/eplbib}
\bibliography{foo}

\end{document}